\documentclass[onecolumn,amsmath,amssymb]{qm2008_abs}
\usepackage{graphicx}
\usepackage{dcolumn}
\usepackage{bm}
\begin{document}

\title{{\Large Parton rearrangement and fast thermalization 
in heavy-ion collisions at RHIC and LHC energies}}

\bigskip
\bigskip
\author{\large G.~Burau}
\email{burau@th.physik.uni-frankfurt.de}
\affiliation{Institut f{\"u}r Theoretische Physik, Johann Wolfgang Goethe-Universit{\"a}t, Max-von-Laue-Str. 1, D-60438 Frankfurt am Main, Germany}
\author{\large J.~Bleibel}
\email{bleibl@tphys.physik.uni-tuebingen.de}
\affiliation{Institut f{\"u}r Theoretische Physik, Eberhard Karls Universit{\"a}t, Auf der Morgenstelle 14, D-72076 T{\"u}bingen, Germany}
\author{\large C.~Fuchs}
\email{christian.fuchs@uni-tuebingen.de}
\affiliation{Institut f{\"u}r Theoretische Physik, Eberhard Karls Universit{\"a}t, Auf der Morgenstelle 14, D-72076 T{\"u}bingen, Germany}
\bigskip
\bigskip

\begin{abstract}
\leftskip1.0cm
\rightskip1.0cm
The implications of parton rearrangement processes on the dynamics of ultra-relativistic 
heavy-ion collisions have been investigated. A microscopic transport approach, namely the 
quark gluon string model (QGSM) which has been extended for a locally density-dependent 
partonic rearrangement and fusion procedure served as the tool for this investigations. 
The model emulates effectively the dynamics of a strongly coupled quark plasma and final 
hadronic interactions. Main QGSM results on anisotropic flow components $v_1$ and $v_2$ 
at top RHIC energy are compiled. Predictions for the pseudorapidity dependence of directed 
and elliptic flow in Pb+Pb collisions under LHC conditions are presented.
\end{abstract}

\maketitle

\section{Introduction}
\label{intro}

Calculations within Quantum Chromodynamics (QCD) realized on the lattice predict a new 
state of matter consisting of deconfined partons at energy densities above $\approx 1~{\rm GeV/fm^3}$ 
and temperatures above $\approx 170~{\rm MeV}$ \cite{qm2006,Karsch:2006sm,Aoki:2006br}. 
Experimental investigations of Au+Au reactions at center of mass energies up to 
$\sqrt{s_{NN}} = 200~{\rm GeV}$ at the Relativistic Heavy Ion Collider (RHIC) led to observations 
of large elliptic flow \cite{Ackermann:2000tr,Park:2001gm,Manly:2002uq,Adler:2003kt,Back:2004mh,Back:2005pc}, 
which is one of the key signals justifying a strongly coupled quark-gluon plasma (sQGP) 
formed in the early phase of such high-energetic heavy-ion collisions (see \cite{qm2006} 
and references therein for a detailed discussion). A system of this kind, which does not 
behave like a weakly interacting gas of quarks and gluons, implies large pressure gradients 
and short equilibration times \cite{Heinz:2001xi,Shuryak:2003xe}. Both are necessary 
conditions for the development of large elliptic flow. The description of various measured 
anisotropic flow patterns by hydrodynamic approaches (see, e.g., \cite{Hirano:2005xf}) 
together with the scaling behavior of the elliptic flow with the number of constituents 
(see, e.g., \cite{Abelev:2007qg,Bass:2006ib}) well explained by parton recombination/coalescence 
approaches \cite{Dover:1991zn,Hwa:2002tu,Hwa:2002tu_b,Greco:2003xt,Greco:2003xt_b,Greco:2003xt_c,Molnar:2003ff,Molnar:2003ff_b,Fries:2003kq,Fries:2003kq_b}, give hints towards the partonic nature of the created matter 
which equilibrates very fast.

In this paper, the main anisotropic flow results obtained with a microscopic transport approach, 
namely the Quark-Gluon String Model (QGSM) extended for a dynamical mechanism which allows for 
quark rearrangement and fusion processes during the very dense stages of ultra-relativistic 
heavy-ion reactions, are summarized \cite{Bleibel:2006xx,Bleibel:2007se}. After a brief 
account of the approach and its conceptual ideas in Sect. \ref{qgsm+reco}, the implications 
on kinetic equilibration and anisotropic flow components ($v_1$, $v_2$) at top RHIC energy 
are compiled in Sect. \ref{RHIC+LHC-flow}, followed by QGSM predictions for the pseudorapidity 
dependence of $v_1$ and $v_2$ in Pb+Pb collisions under LHC conditions.

\section{Quark Gluon String Model with quark rearrangement}
\label{qgsm+reco}

\begin{figure}[h]
\includegraphics[scale=0.28]{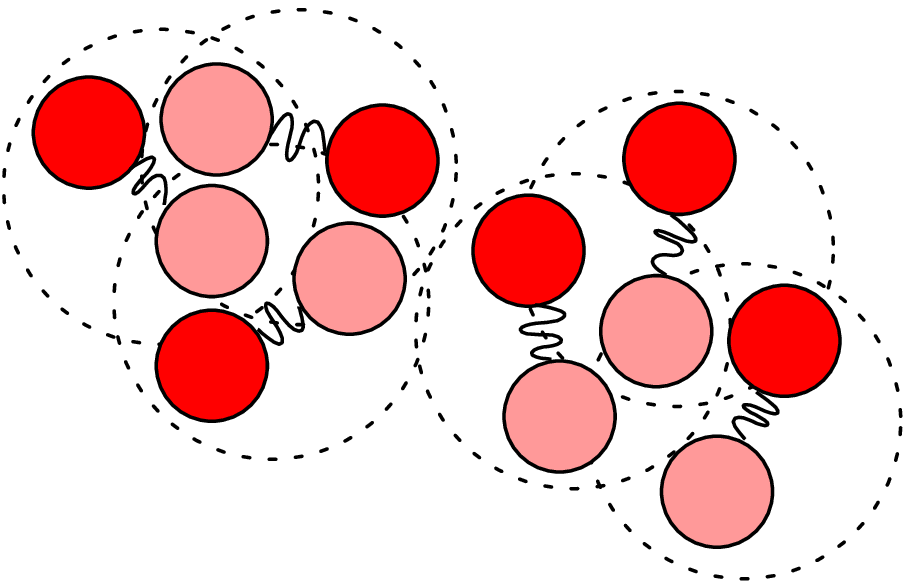}
\hspace{25pt}
\includegraphics[scale=0.28]{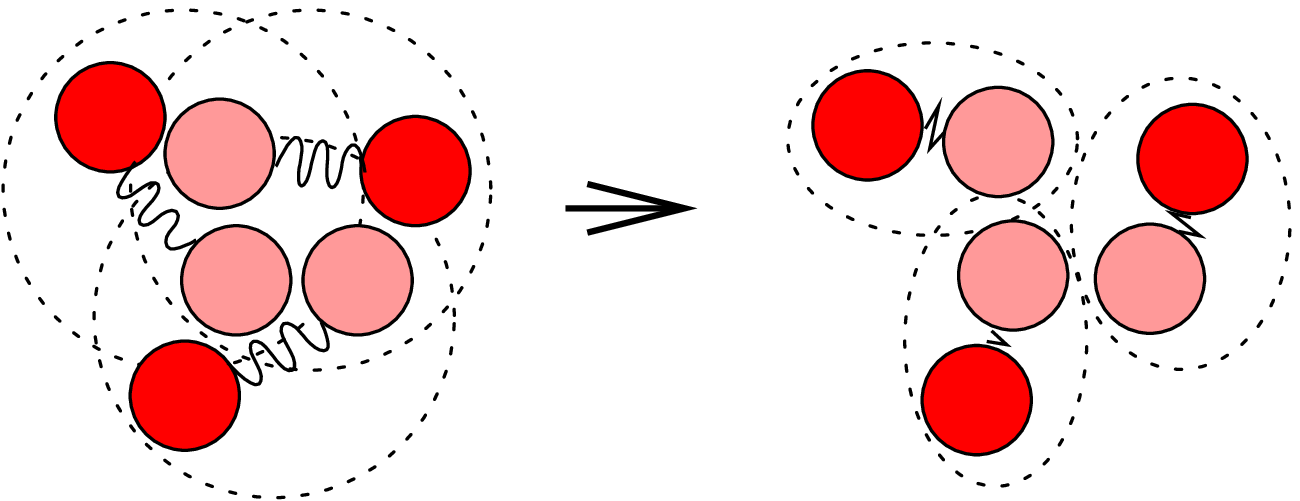}
\hspace{25pt}
\includegraphics[scale=0.45]{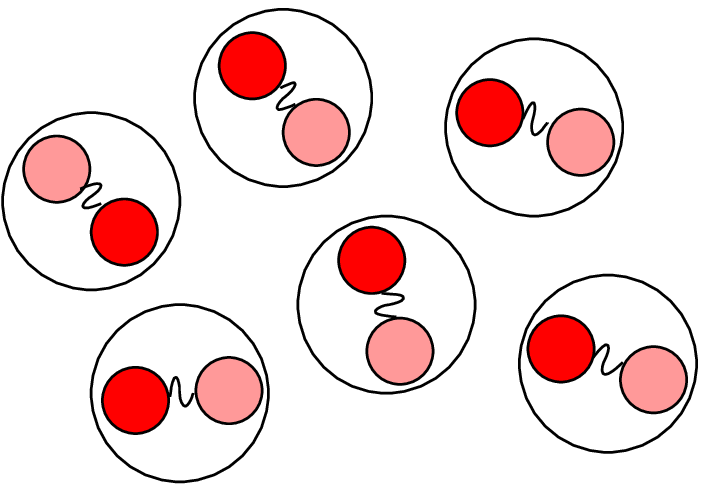}
\caption{Schematical view on the conception of the quark rearrangement mechanism 
implemented in the QGSM. Red and pink ``blobs'' illustrate quarks and antiquarks, 
respectively. Dashed black circles indicate strongly correlated states, whereas 
solid black circles denote hadrons (more details in Sect. \ref{qgsm+reco}).
\label{fig:recoconcept}}
\end{figure}

The standard version of the QGSM -- a microscopic string cascade transport approach 
using Monte-Carlo simulation techniques in order to study theoretically heavy-ion 
collisions at high energies -- describes particle production by excitation and decay 
of classical strings with (anti)quarks or (anti)diquarks at their ends, like other 
transport approaches of this kind. A more detailed description of the standard QGSM 
can be found in \cite{QGSM0,QGSM0_b,QGSM1c}. Since in the very dense stages of a 
heavy-ion reaction ``hadrons'' overlap each other and consequently are rather strongly 
correlated partonic states than real bound states (schematically depicted by the left 
picture of Fig. \ref{fig:recoconcept}), locally density-dependent partonic rearrangement 
and fusion processes have been implemented in the QGSM. 
Above a critical local particle density, ``hadronic correlations'' are decomposed 
into their constituent partons which afterwards rearrange among themselves into 
new hadron-like correlations. This rearrangement process is sketchily displayed 
in the middle of Fig. \ref{fig:recoconcept}. A quark-antiquark pair of same flavor 
may additionally annihilate during the rearrangement process. Thus, a $3 \to 2$ 
reaction is effectively implemented, which is a new feature within a string cascade 
approach. Rearrangement and fusion processes become unimportant when the system 
increasingly thins out due to ongoing expansion. Finally, hadrons are the relevant 
degrees of freedom as shown in the right picture of Fig. \ref{fig:recoconcept}. 
In this spirit the extended QGSM emulates the dynamics of a sQGP and final hadronic 
interactions from a microscopical point of view. The concept and procedure of the 
quark rearrangement and fusion mechanism and its implementation into the QGSM are 
described in detail in Refs. \cite{Bleibel:2006xx,Bleibel:2007se}.

\section{Kinetic equilibration and anisotropic flow at RHIC and LHC}
\label{RHIC+LHC-flow}

\begin{figure}[htb]
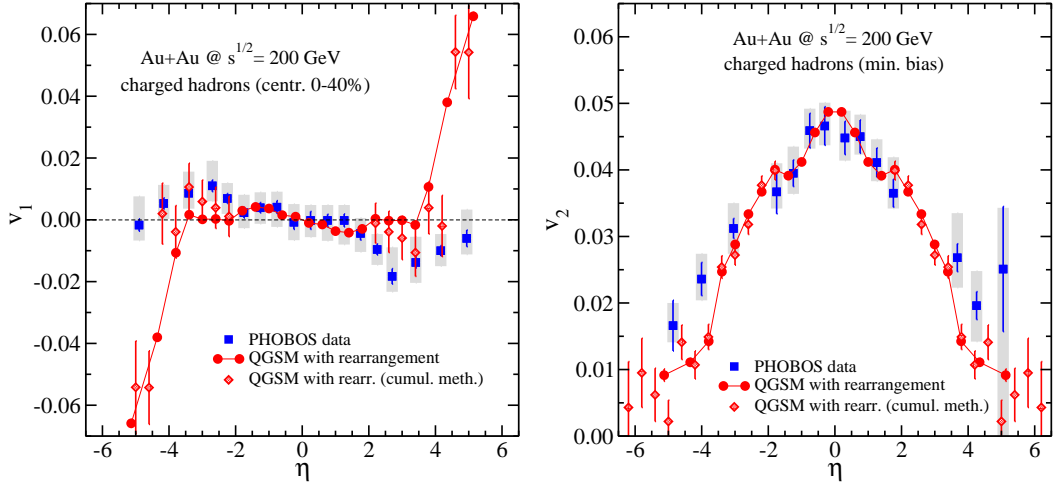

\includegraphics[scale=0.425]{v1etaAuAu200GeV_hchar_reco.eps}
\hspace{3pt}
\includegraphics[scale=0.425]{v2etaAuAu200GeV_hchar_reco.eps}
\caption{Directed flow (left panel) and elliptic flow (right panel) of charged hadrons 
as function of pseudorapidity $\eta$. The results from simulations of Au+Au collisions at 
$\sqrt{s_{NN}} = 200~{\rm GeV}$ with the QGSM including quark rearrangement (filled circles) 
\cite{Bleibel:2006xx} together with corresponding results from an analysis applying a cumulant 
method (diamonds) are shown in comparison to PHOBOS data (filled squares) \cite{Back:2005pc}. 
Systematic errors of the experimental data are indicated by gray boxes, statistical errors by bars.
\label{fig:v1v2etaRHIC}}
\end{figure}

It has been demonstrated in \cite{Bleibel:2006xx} that dynamical parton rearrangement 
which occurs in the very dense medium created in ultra-relativistic heavy-ion reactions 
during the early stages drives the system to fast kinetic equilibrium. The time dependence 
of the equilibration ratio $R_{\rm LE}$ in the overlap zone of Au+Au collisions simulated by 
using QGSM with and without quark rearrangement has been extensively discussed in references 
\cite{Bleibel:2006xx,Bleibel:2005gx}. Quark rearrangement processes reduce the local 
equilibration time of the system roughly by a factor of 5, i.e., from $\approx 10~{\rm fm/c}$ 
to $\approx 2~{\rm fm/c}$. Moreover, the rearrangement mechanism improves significantly 
on the theoretical description of measured directed and elliptic flow, i.e., $v_1$ and $v_2$ 
distributions and their pseudorapidity dependence in Au+Au collisions at top RHIC energy 
of $\sqrt{s_{\rm NN}} = 200~{\rm GeV}$ as shown in Fig. \ref{fig:v1v2etaRHIC}. 
In particular the shape of $v_2(\eta)$ is found to be closely related to fast thermalization. 
A detailed comparison and discussion of the simulation results on $v_1$ and $v_2$ using the 
QGSM with and without implementation of quark rearrangement and fusion processes can be found 
in \cite{Bleibel:2006xx,Burau:2004ev,Zabrodin:2005pd,Bravina:2004td}. The centrality dependence 
of $v_2(\eta)$ has been discussed in \cite{Bleibel:2006xx,Burau:2004ev}. 
A fair description of this observable has been achieved so far only by a hydro-cascade 
hybrid model \cite{Hirano:2005xf}, the partonic rearrangement ansatz within the microscopic 
QGSM \cite{Bleibel:2006xx}, and recently by a parton cascade approach \cite{Xu:2007jv}.

\begin{figure}[h]
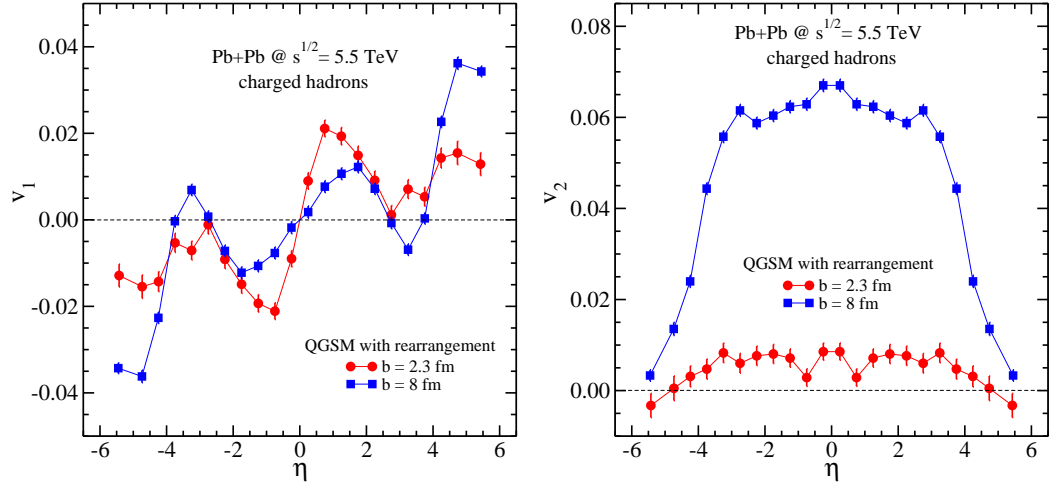

\includegraphics[scale=0.425]{v1etaPbPb5TeV_hchar_reco.eps}
\hspace{3pt}
\includegraphics[scale=0.425]{v2etaPbPb5TeV_hchar_reco.eps}
\caption{The same as in Fig. \ref{fig:v1v2etaRHIC}, but for Pb+Pb collisions at 
$\sqrt{s_{\rm NN}} = 5.5~{\rm TeV}$. Predictions by QGSM simulations including 
quark rearrangement and fusion processes are depicted for impact parameters 
$b = 2.3~{\rm fm}$ (filled circles) and $b = 8~{\rm fm}$ (filled squares), 
respectively \cite{Bleibel:2007se}. Error bars denote statistical uncertainties.
\label{fig:v1v2etaLHC}}
\end{figure}

Predictions by the improved QGSM for the pseudorapidity dependence of the azimuthal anisotropy 
parameters $v_1$ and $v_2$ of charged hadrons produced in central and semiperipheral Pb+Pb 
reactions at a LHC energy of $\sqrt{s_{\rm NN}} = 5.5~{\rm TeV}$ are compiled in 
Fig. \ref{fig:v1v2etaLHC} \cite{Bleibel:2007se}.

The simulation results, e.g., an increasing magnitude of $v_2$ around midrapidity under LHC conditions 
by about 10-20\% compared to the RHIC results, suggest in comparison with results of other transport 
approaches and hydrodynamical calculations that the hydrodynamical limit will be reached at this collision 
energy. A more detailed discussion can be found in \cite{Bleibel:2007se}.

\section{Summary and conclusions}

The implications of locally density dependent quark rearrangement and fusion processes on the 
collision dynamics have been compiled. In addition to a much faster kinetic equilibration 
achieved in simulated ultra-relativistic heavy-ion reactions, the magnitude and shape of $v_1(\eta)$ 
and $v_2(\eta)$ within the rearrangement scenario are in nice agreement with the experimental 
data of the RHIC program. From a microscopical point of view, this is a strong indication for 
a partonic medium with short mean free path and accordingly rather low viscosity created in 
such collisions at early times. The predicted increase of the elliptic flow at 
$\sqrt{s_{\rm NN}} = 5.5~{\rm TeV}$ is in line with conclusions that the hydrodynamical limit 
will be reached under LHC conditions.\\

\end{document}